\documentclass{aa}
\usepackage{graphicx}
\usepackage{txfonts}
\usepackage{natbib}

\begin{document}

\def\be{\begin{equation}}
\def\ee{\end{equation}}
\def\bea{\begin{eqnarray}}
\def\eea{\end{eqnarray}}
\def\c{\cite}
\def\nn{\nonumber}
\def\mcr{{{\rm M_{cr}}}}
\def\xo{{X_{o}}}
\def\dm{\Delta {\rm M}}
\def\ms{{\rm M_{\odot}}}
\def\mb{m_{B}}
\def\bo{B_{o}}
\def\cxo{C(x_{o})}
\def\rsix{R_{6}}
\def\vr{v_{{\rm r}}}
\def\vro{v_{{\rm ro}}}
\def\vt{v_{\theta}}
\def\po{\ifmmode P_{o} \else $P_{o}$ \fi}

\def\et{ {\it et al.}}
\def\la{ \langle}
\def\ra{ \rangle}
\def\ov{ \over}
\def\ep{ \epsilon}

\def\ep{\epsilon}
\def\th{\theta}
\def\ga{\gamma}
\def\Ga{\Gamma}
\def\la{\lambda}
\def\si{\sigma}
\def\al{\alpha}
\def\pa{\partial}
\def\de{\delta}
\def\De{\Delta}
\def\rsr{{r_{s}\over r}}
\def\rmo{{\rm R_{M0}}}
\def\rrm{{R_{{\rm M}}}}
\def\rra{{R_{{\rm A}}}}

\def\mdot{\ifmmode \dot M \else $\dot M$\fi}    
\def\mxd{\ifmmode \dot {M}_{x} \else $\dot {M}_{x}$\fi}
\def\med{\ifmmode \dot {M}_{Edd} \else $\dot {M}_{Edd}$\fi}
\def\bff{\ifmmode B_{{\rm f}} \else $B_{{\rm f}}$\fi}

\def\apj{\ifmmode ApJ \else ApJ \fi}    
\def\apjl{\ifmmode  ApJ \else ApJ \fi}    %
\def\aap{\ifmmode A\&A \else A\&A\fi}    %
\def\mnras{\ifmmode MNRAS \else MNRAS \fi}    %
\def\nat{\ifmmode Nature \else Nature \fi}
\def\prl{\ifmmode Phys. Rev. Lett. \else Phys. Rev. Lett.\fi}
\def\prd{\ifmmode Phys. Rev. D. \else Phys. Rev. D.\fi}

\def\ms{\ifmmode M_{\odot} \else $M_{\odot}$\fi}    
\def\na{\ifmmode \nu_{A} \else $\nu_{A}$\fi}    
\def\nk{\ifmmode \nu_{K} \else $\nu_{K}$\fi}    
\def\ns{\ifmmode \nu_{{\rm s}} \else $\nu_{{\rm s}}$\fi}
\def\no{\ifmmode \nu_{1} \else $\nu_{1}$\fi}    
\def\nt{\ifmmode \nu_{2} \else $\nu_{2}$\fi}    
\def\ntk{\ifmmode \nu_{2k} \else $\nu_{2k}$\fi}    
\def\dnmax{\ifmmode \Delta \nu_{max} \else $\Delta \nu_{2max}$\fi}
\def\ntmax{\ifmmode \nu_{2max} \else $\nu_{2max}$\fi}    
\def\nomax{\ifmmode \nu_{1max} \else $\nu_{1max}$\fi}    
\def\nh{\ifmmode \nu_{\rm HBO} \else $\nu_{\rm HBO}$\fi}    
\def\nqpo{\ifmmode \nu_{QPO} \else $\nu_{QPO}$\fi}    
\def\nz{\ifmmode \nu_{o} \else $\nu_{o}$\fi}    
\def\nht{\ifmmode \nu_{H2} \else $\nu_{H2}$\fi}    
\def\ns{\ifmmode \nu_{s} \else $\nu_{s}$\fi}    
\def\nb{\ifmmode \nu_{{\rm burst}} \else $\nu_{{\rm burst}}$\fi}
\def\nkm{\ifmmode \nu_{km} \else $\nu_{km}$\fi}    
\def\ka{\ifmmode \kappa \else \kappa\fi}    
\def\dn{\ifmmode \Delta\nu \else \Delta\nu\fi}    

\def\rs{\ifmmode {R_{s}} \else $R_{s}$\fi}    
\def\ra{\ifmmode R_{A} \else $R_{A}$\fi}    
\def\rso{\ifmmode R_{S1} \else $R_{S1}$\fi}    
\def\rst{\ifmmode R_{S2} \else $R_{S2}$\fi}    
\def\rmm{\ifmmode R_{M} \else $R_{M}$\fi}    
\def\rco{\ifmmode R_{co} \else $R_{co}$\fi}    
\def\ris{\ifmmode {R}_{{\rm ISCO}} \else $ {\rm R}_{{\rm ISCO}} $\fi}
\def\rsix{\ifmmode {R_{6}} \else $R_{6}$\fi}
\def\rinfty{\ifmmode {R_{\infty}} \else $R_{\infty}$\fi}
\def\rinfsix{\ifmmode {R_{\infty6}} \else $R_{\infty6}$\fi}

\def\rxj{\ifmmode {RX J1856.5-3754} \else RX J1856.5-3754\fi}
\def\1739{\ifmmode {XTE  J1739-285} \else XTE  J1739-285\fi}
\def\exo{\ifmmode {EXO 0748-676} \else EXO 0748-676\fi}

\title{Spin Period Evolution of Recycled
 Pulsar in Accreting Binary}

\author{J. Wang
          \inst{1}
          \and
          C.M.  Zhang
          \inst{1}
          \and
          Y,H,  Zhao
          \inst{1}
          \and
          Y. Kojima
          \inst{2}
          \and
          H. X.  Yin
          \inst{3}
          \and
          L.M. Song
          \inst{4}
          }
\institute{National Astronomical Observatories, Chinese Academy
of Sciences, Beijing 100012, China, \\
\email{zhangcm@bao.ac.cn} \and Department of Physics, Hiroshima
University, Higashi-Hiroshima 739-8526, Japan\\ \and School of Space
Science and Physics, Shandong University, Weihai 264209, China\\
\and Institute of High Energy Physics, Chinese Academy of Sciences,
Beijing 100049, P. R. China}

\abstract {We investigate the spin-period evolutions of recycled
pulsars in binary accreting systems. Taking both the accretion
induced field decay and spin-up into consideration, we calculate
their spin-period evolutions influenced by the initial
magnetic-field strengths, initial spin-periods and accretion rates,
respectively. The results indicate that the minimum spin-period (or
maximum spin frequency) of millisecond pulsar (MSP) is independent
of the initial conditions and accretion rate when the neutron star
(NS) accretes $\sim> 0.2\ms$. The accretion torque with the fastness
parameter and gravitational wave (GW) radiation torque may be
responsible for the formation of the minimum spin-period (maximum
spin frequency). The fastest spin frequency (716 Hz) of MSP can be
inferred to associate with a critical fastness parameter about
$\omega_{c}=0.55$. Furthermore, the comparisons with the
observational data are presented in the field-period ($B-P$) diagram. 
}

\maketitle

\section{Introduction}

\begin{figure*}
\centering $\begin{array}{c@{\hspace{0.2in}}c}
\multicolumn{1}{l}{\mbox{}} &
\multicolumn{1}{l}{\mbox{}} \\
\includegraphics[width=8cm]{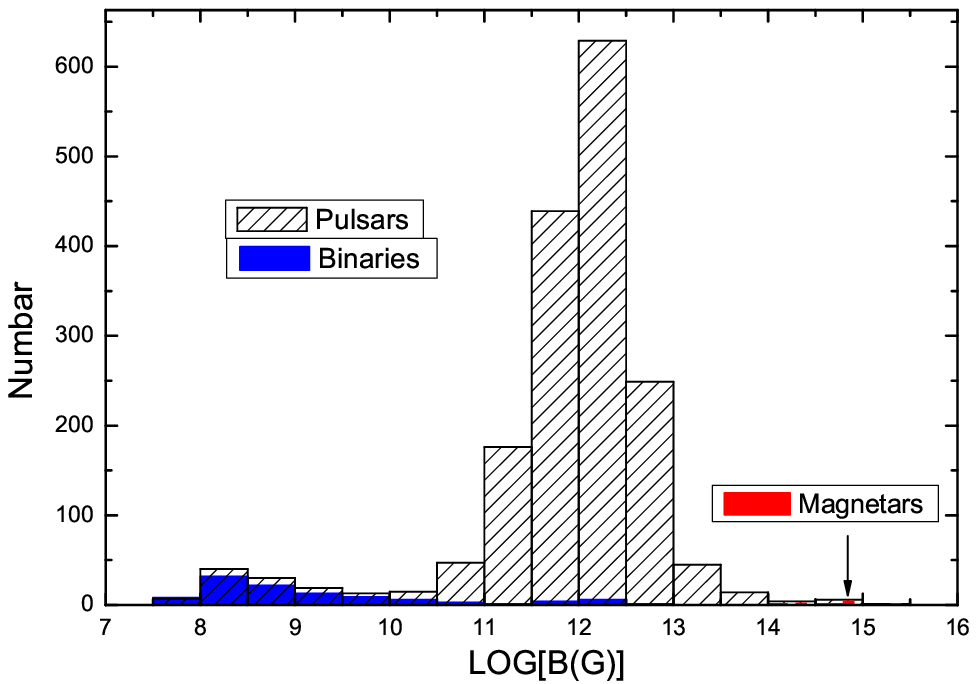} &\includegraphics[width=8cm]{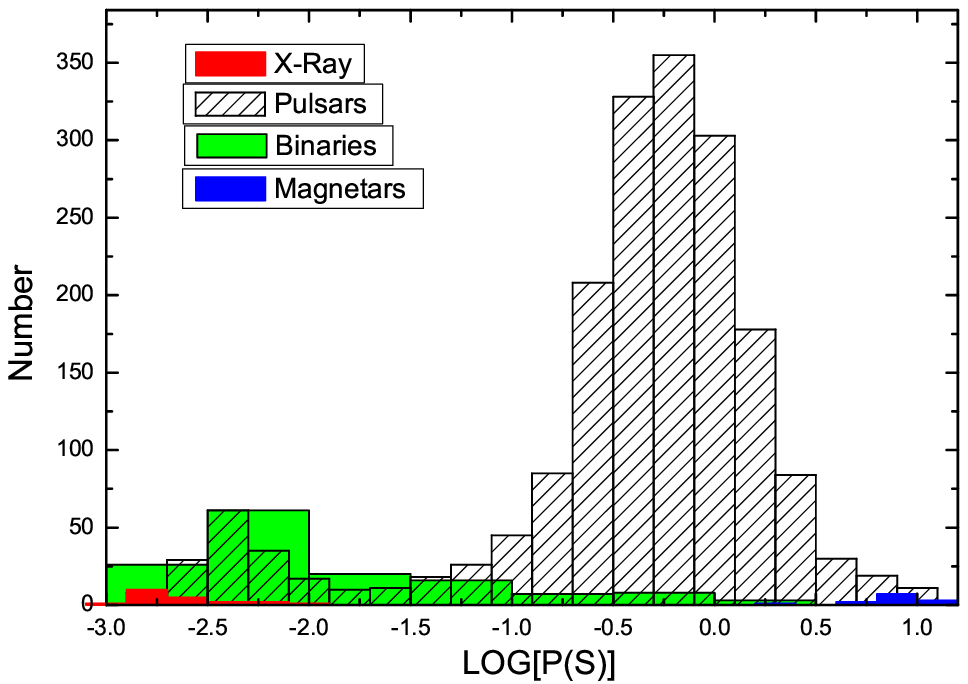} \\
\end{array}$
\caption{Distribution of magnetic-field and spin-period for 1864
pulsars (data  from the ATNF pulsar catalogue, see Manchester et al.
2005). Here, 141 binary pulsars and 13 magnetars (Kouveliotou 2003)
are separately identified for the statistics of magnetic-field and
spin-period. The 24 accretion millisecond X-ray pulsars discovered
by RXTE (see e.g. Wijnands 2005; Lamb \& Yu 2005; Yin et al. 2007)
are put in the spin-period histogram.}\label{his}
\end{figure*}

Neutron stars (NSs) are usually detected as either normal pulsars
(single or in a binary) with magnetic-field $B \sim 10^{12}$ G and
spin-period $P \sim 0.5$ s, or millisecond pulsars (MSPs) (half in
binary) with $B \sim 10^{8.5}$ G and $P \sim 20$ ms (e.g.
Bhattacharya \& van den Heuvel 1991; Lorimer 2008). The spin-periods
and surface magnetic-fields assigned to these two different systems
are found to span different ranges of $B-P$ values (Hobbs \&
Manchester 2004; Manchester et al. 2005; Lorimer 2008), and the
updated statistical distributions of $B$ and $P$ are plotted in
Fig.~\ref{his} (data from ATNF pulsar catalogue). In this figure, we
notice that the distributions of both $B$ and $P$ are almost
bimodal, as firstly suggested by Camilo et al. (1994), with a
dichotomy between "normal" pulsars ($P \sim 0.1$ s - $10$ s and $B
\sim 10^{11}$ G - $10^{13}$ G) and MSPs ($P \sim 1$ ms - $20$ ms and
$B \sim 10^{8}$ G - $10^{9}$ G). This abnormal phenomenon has led to
the idea that radio pulsars in binary systems have been "recycled",
i.e. they have been spun up due to mass accretion during the phase
of mass exchange in binaries (Alpar et al. 1982; Radhakrishnan \&
Srinivasan 1982; Bhattacharya \& Srinivasan  1995). The weak
magnetic-field of MSP strongly supports the idea that it decays in
the binary accretion phase (e.g. Taam \& van den Heuvel 1986; van
den Heuvel 2004). The currently widely accepted view about this
phenomenon is the accretion-induced evolution, i.e., NSs accreting
materials from their low-mass binary companions are spun up by the
angular momentum carried by the accreted materials, while the
magnetic-field decays (Taam \& van den Heuvel 1986; Bhattacharya \&
van den Heuvel 1991; van den Heuvel 2004). The X-ray binaries are
the evolutionary precursors to these "recycled"  MSPs. It is evident
that $B$ and $P$ of X-Ray pulsars and recycled pulsars are
correlated with the duration of both the accretion phase and the
total amount of matter accreted (Taam \& van den Heuvel 1986;
Shibazaki et al. 1989). If the NS accretes a small quantity of mass
from its companion, e.g. $\sim 0.001\ms - 0.01\ms$, a recycled
pulsar with the mildly weak field and short spin-period ($B \sim
10^{10}$ G, $P \sim 50$ ms) will be formed (e.g. Francischelli,
Wijers \& Brown 2002), like PSR 1913+16 and PSR J0737-3039 (Lyne et
al. 2004). The direct evidence for this recycled idea has been found
in low mass X-ray binary (LMXB) with the accretion  millisecond
X-ray pulsar (AMXP), e.g. Sax J 1808.4-3658 (Wijnands \& Klis 1998),
and in observing the transition link from an X-ray binary to a radio
pulsar PSR J1023+0038 (Archibald et al. 2009).

At the end of the accretion phase of NS/LMXB (accreted mass $\sim
> 0.2 \ms$), the NS magnetic-field may arrive at a bottom value about
$10^{8}$ G and spin-period may reach a minimum about millisecond
(Cheng \& Zhang 1998, 2000), remaining a MSP (van den Heuvel \&
Bitzaraki 1995ab). Since then, spin-down due to dipole radiation has
been conceived as the "standard evolution" of MSP (Urpin \& Konenkov
1997; Urpin, Geppert \& Konenkov 1997, 1998). The switch-on of a
radio pulsar could exist, whose radiation pressure is capable of
ejecting the accreted matter out of the system and prevent any
further accretion. This 'radio ejection' phase in the binary will
account for the formation of a MSP (Burderi et al. 2002ab). However,
the question whether all observed MSPs could be produced within this
recycled framework has not been quantitatively addressed until now.
The accretion induced collapse (AIC) of a white dwarf as an
alternative mechanism for the formation of MSP has been studied
recently (Ferrario \& Wickramasinghe 2007ab; Hurley et al. 2009;
Wickramasinghe et al. 2009). Recently, another idea for MSP
formation is proposed, which claims that these objects are born with
low magnetic fields (Halpern \& Gotthelf 2010).

Then the mechanisms for the accretion-driven field decay and spin-up
in binary NSs have been suggested by many researchers in various
proposals. The accretion flow and thermal effects speed-up Ohmic
dissipation of NS crust currents, which accounts for the field decay
and spin evolution (Geppert \& Urpin  1994; Geppert, Page \& Zannias
1999; Urpin \& Geppert  1995; Romani 1990). The interactions between
the accreted matter and magnetic-field in the course of accretion
phase are studied to account for the field decay (Melatos \& Phinney
2001; Payne \& Melatos  2004; Konar \& Bhattcharya 1999; Konar \&
Choudhury  2004; Lovelace, Romanova \& Bisnovatyi-Kogan 2005). Not
only is the total amount of accreted mass considered to influence
the final field strength and minimum spin-period (van den Heuvel
1995; van den Heuvel \& Bitzaraki 1994, 1995ab), but also the
influences of accretion rate may be a factor (Wijers 1997; Cumming
et al. 2001; Cumming 2005; Zhang \& Kojima  2006).

Taking the accretion-induced field decay into consideration, we
investigate the formation of the high spin frequency (short
spin-period) for a MSP, and the influences on it by the initial
conditions (e.g. initial spin-period and initial field) and
accretion rate. The spin-up torque and field evolution make their
spin-period and field distributions in $B-P$ diagram to change from
the regions of $B \sim 10^{11-13} $ G and $P \sim 1$ s - $100$ s to
that of $B \sim 10^{8-9} $ G and $ P \sim 1$ ms - $20$ ms as
observed. This paper is organized as follows. In Section 2, we
review the main equations that dominate the $B-P$ relation in the
accretion phase. The effects by the initial conditions and accretion
rate on the spin-period are presented in Section 3. Section 4
contains the summary and discussion.

\section{Field Versus Period Relation}

\subsection{Description for the Model}

As declaimed by the accretion induced magnetic field decay model
(Zhang \& Kojima 2006), if the magnetic field is sufficiently
strong, e.g. $\sim 10^{12}$ G, and the  star spin very slowly
initially, the accreted matter will be channeled onto the polar
patches by the field lines, where the compressed accreted matter
causes  the expansion of magnetic polar zone in two directions,
downward and equatorward, which makes the magnetic flux in the polar
zone  diluted. With accretion, the field decays and magnetosphere
shrinks until the magnetosphere reaches the surface of NS, where a
bottom field is obtained to be about $\sim 10^{8}$ G.

Based on the above model, the accretion-induced field evolution is
obtained analytically with the initial field $B(t=0) = B_0$ (Zhang
\& Kojima 2006),
\begin{equation}
\label{bt} B = \frac{B_f}{(1 - [C~exp(-y)-1]^2)^{\frac{7}{4}}}.
\end{equation}
Here, we define the parameters as follows, $y = \frac{2 \Delta
M}{7M_{cr}}$, the accreted mass $\Delta M = \dot{M} t$, the crust
mass $M_{cr} \sim 0.2 M_\odot$, $C = 1 + \sqrt{1-x_0^2} \sim 2$ with
$x_0^{2} = (\frac{B_f}{B_0})^{4/7}$. $B_f$ is the bottom
magnetic-field which is defined by the NS magnetosphere radius
matching the stellar radius, i.e., $R_M(B_f) = R$. $\rrm$ is defined
as $R_M = \phi R_A$ where $Alfv\acute{e}n$ radius $R_A$ (Elsner \&
Lamb 1977; Ghosh \& Lamb 1977) reads, \bea R_A =3.2\times 10^8 ({\rm
cm}) \dot{M}^{-2/7}_{17} \mu^{4/7}_{30} m^{-1/7}\,\,\, ,
\eea The model dependent parameter $\phi$ is about 0.5 (Ghosh \&
Lamb 1979b hereafter GL; Shapiro \& Teukolsky 1983; Frank et al.
2002). $\dot{M}_{17}$ is the accretion rate in units of $10^{17}$
g/s. $\mu _{30}$ is the magnetic moment in units of $10^{30} {~}
{\rm G cm^3}$. The mass $m = M/M_{\odot}$ is in the unit of solar
mass. Using the relation $R_M(B_f) = R$, we can obtain the bottom
field,
\begin{equation}
B_f=1.32\times10^8(G)(\frac{\dot{M}}{\dot{M}_{18}})^{\frac{1}{2}}
m^{\frac{1}{4}}R_6^{-\frac{5}{4}}\phi^{-\frac{7}{4}},
\end{equation}
where $\dot{M}_{18} = \dot{M}/10^{18} g/s$ and $\rsix = R/10^6 cm$.
As declaimed by Inogamov \& Sunyaev (1999), in the accretion
process, a boundary layer forms between the innermost disk and
magnetosphere due to the transition for rotating velocity of plasma
from Keplerian to the spin velocity of NS.  When the accretion rate
increases, this layer may spread (Inogamov \& Sunyaev 1999), and as
a result   more matter is accreted to the polar gap and diffuses to
the entire surface of the NS.  {\bf This effect will influence on
the field decay efficiency of the  model by Zhang \& Kojima (2004),
since some spreading matter has little contribution to the field
lines dragging of the polar patches. Moreover, such a spreading will
expand the polar cap area, which makes the polar cap to occupy the
entire  NS surface  while  the magnetosphere  does not reach  the NS
radius.  The bottom field of NS is determined by the condition that
 the magnetosphere equals the NS radius (Zhang \& Kojima 2004).  If
 there is no field decay while the magnetosphere radius is bigger
 than NS radius, then the modified  bottom field  by considering the spreading
 of the polar cap  will be bigger than the one  obtained for the
ideal frozen plasma in polar cap zone. Namely, the magnetic field of
recycled NS is slight bigger than the ideal value of the model by
Zhang \& Kojima (2004).

}

 To study the spin evolution of NS in accretion phase, we employ
the formula for accretion-induced spin-up given by GL,
\begin{eqnarray}
-\dot{P}&=&5.8 \times 10^{-5}[(\frac{M}{M_\odot})^{-\frac{3}{7}}
R^{\frac{12}{7}}_6 I_{45}^{-1}]\nonumber\\~& & \times
B^{\frac{2}{7}}_{12}
(PL^{\frac{3}{7}}_{37})^2n(\omega_s) {\empty ~ ~ }  syr^{-1},\label{pdot}
\end{eqnarray}
where we define the parameters, the surface field $B_{12} =
B/10^{12} G$, the moment of inertia $I_{45} = I/10^{45} gcm^2$, the
X-ray brightness ($L = GM\dot{M}/R$) $L_{37}$ in units of $10^{37}
erg/s$, respectively. The dimensionless parameter $n(\omega_s)$ is
the fastness parameter which is explained in the following
subsection.

\subsection{Fastness Parameter}

For a slowly rotating magnetic NS, the rotation of the star couples
to the orbital motion of the disk matter at the Alfv\'{e}n radius
when accretion flow is approximately radial. So the limit of angular
velocity ($\Omega_s = 2\pi/P$) for a NS can reach the Keplerian
velocity ($\Omega_k(R_{M}) = \sqrt{GM/R_M^3}$). In general, the spin
velocity is always less than the orbital velocity, i.e. $\Omega_s
\ll\Omega_k(R_{M}) = \sqrt{GM/R_M^3}$. Therefore, the relative
importance of stellar rotation can be described by the ratio
parameter of the angular velocities (Elsner \& Lamb 1977; Ghosh \&
Lamb 1977; Li \& Wang 1996, 1999; Shapiro \& Teukolsky 1983), \be
\omega_s\equiv\frac{\Omega_s}{\Omega_k(R_M)} =
1.35[(\frac{M}{M_\odot})^{-2/7}R^{15/7}_6] B^{6/7}_{12} P^{-1}
L^{-3/7}_{37} \,\label{omega}, \ee which plays a significant role in
our entire understanding of accretion to the rotating magnetic NSs.

To calculate the accretion torque which acts on a magnetic NS
accreting matter from a disk, GL introduced a dimensionless
accretion torque $n(\omega_{s})$ depending primarily on the fastness
parameter. A simple expression for $n(\omega_{s})$ is given by GL,
\be n(\omega_{s}) = 1.4\times \left(
\frac{1-\omega_{s}/\omega_{c}}{1-\omega_{s}} \right), \label{ns}\ee
where $\omega_c$ is the critical value. For a slowly rotating star
$(\omega_{s} \ll 1 )$, GL found that the dimensionless function
$n(\omega_{s})$ decreases with $\omega_{s}$ and become negative for
$\omega_{s} > \omega_{c}$ if $n(\omega_{s}) \sim 1.4$. In addition,
GL also stressed $\omega_{c} \sim 0.35$ for their model. However,
the subsequent work (Ghosh \& Lamb 1992) indicates that $\omega_{c}$
is unlikely to be less than $0.2$, but it could be as large as
$0.9$. By the aid of this dimensionless parameter, we can establish
the relation between the total torque on the star and the torque
communicated to the star by the magnetic-field lines that thread the
inner transition zone.

In the following subsections, we study the spin evolution of NS in a
binary system, where the influence by fastness is included.

\section{Spin Evolution of Accreting NS}

In this section, we investigate the spin-period evolution of
accreting NSs by solving equations (\ref{bt}), (\ref{pdot}),
(\ref{omega}) and (\ref{ns}) with different initial conditions and
accretion rates, where we set the usual parameters for NS, e.g. $m =
1.4$, $\rsix = 1.5$, $\phi = 0.5$ and $\mcr = 0.2 \ms$.

\subsection{Influence by Initial Conditions}

\begin{figure}
\includegraphics[width=8cm]{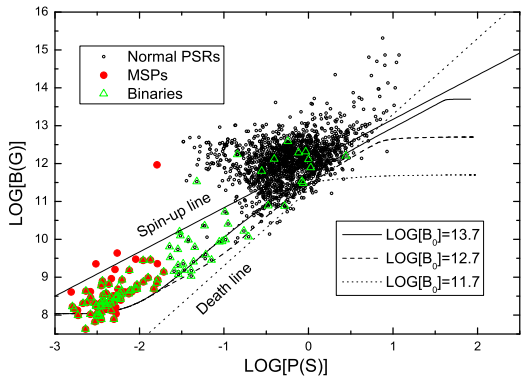}
\caption{The diagram of the magnetic-field versus spin-period for
pulsars. The spin-up line represents the minimum spin-period to
which such a spin-up may proceed in an Eddington-limited accretion,
while the "death-line" corresponds to a polar cap voltage below
which the pulsar activity is likely to switch off (Bhattacharya \&
van den Heuvel 1991). The evolutionary tracks in $B-P$ diagram are
plotted with different initial magnetic-field strengths, but with
same initial spin-period ($\po=100$ s) and same accretion rate
$\dot{M_{17}}$;  The solid, dash and dot lines denote the numerical
solutions for equations (\ref{bt}), (\ref{pdot}), (\ref{omega}) and
(\ref{ns}) with initial field $\bo = 5 \times 10^{13} $ G, 5 $
\times 10^{12} $ G and 5 $\times 10^{11} $ G,
respectively.}\label{B0}
\end{figure}

\begin{figure}
\includegraphics[width=8cm]{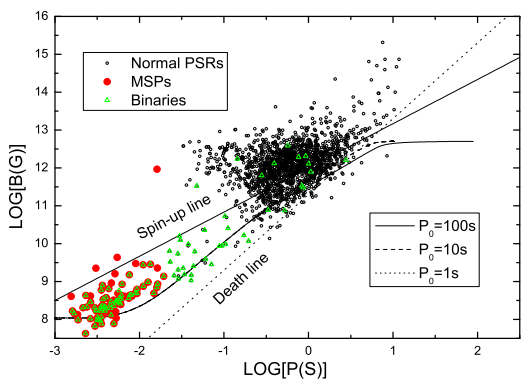}
\caption{The same meaning as Fig. \ref{B0} with different initial
spin-periods ($\po = 1$ s - dot line, $10$ s - dash line, $100$ s -
solid line) at the fixed initial magnetic-field ($\bo=5 \times
10^{12}$ G) and same accretion rate $\dot{M_{17}}$.}\label{P0}
\end{figure}

We consider the spin evolution for a wide range of initial
conditions (initial magnetic-fields and initial spin-periods): (1).
varying the initial magnetic-field $5\times 10^{11}$ G, $5 \times
10^{12}$ G and $5 \times 10^{13}$ G while setting spin-period and
accretion rate as $100$ s and $\dot{M}_{17}$, respectively (Fig.
\ref{B0}); (2). varying the initial spin-period $1$ s, $10$
s~\&~$100$ s with a certain field ($5 \times 10^{12}$ G) and
constant accretion rate
($\dot{M}_{17}$) (Fig. \ref{P0}).

For the samples, we classify them as normal pulsars, MSPs and
binaries. From the figures, we can see that most MSPs are consistent
with the binary pulsars, while only a few binaries fall in the same
region of the normal pulsars that have a magnetic-field $\sim
> 10^{11}$ G.
Recycled pulsars are expected to be found only to the right of the
spin-up line and to the left of the "death-line" (Bhattacharya \&
van den Heuvel 1991), which is shown indeed the case in these
figures (Fig. 2, Fig. 3) for almost all the binaries and MSPs. This
strongly supports the view that the MSPs have been recycled (Alpar
et al. 1982; Radhakrishnan \& Srinivasan 1982). Firstly, with a
small amount of mass transferred, the NS is spun-up from the death
valley with long period and strong field, e.g. the NS in HMXBs ($B
\sim 10^{12}$ G, $P \sim 10$ s) like Her X-1 ($P = 1.24$ s, $B = 3
\times 10^{12}$ G, see van der Meer et al. 2007). With the
long-lived accretion phase the binary accepts  sufficient mass from
its companion  and the NS yield a substantial field decay, as in the
case of MSPs (Bhattacharya \& van den Heuvel 1991), e.g., the
fastest known pulsar PSR J1748-2446 ($B \sim 10^{8}$  G, P = 1.4 ms,
see Hessels et al. 2006).

As shown in Fig.2, three B-P tracks with the different initial field
values begin to follow one track when the spin period evolves to the
hundred millisecond regime, and here the system accretes $\sim 0.01
\ms$. After the system accretes about $\sim$ 0.2 \ms, the field
reaches $\sim 10^{8} $ G -- a bottom value, and the spin period
enters the millisecond regime, which is inferred corresponding to
the condition $\omega_{s} = \omega_{c}$. Namely, the bottom spin
period or maximum  spin frequency can be obtained by \be \nu_{sMax}
= \nu_{k}(\rmm=R)\omega_{c}\;,
 \ee
where $\nu_{k}(\rmm=R) = 1300 (Hz) (M/1.4\ms)^{1/2}(R/15
km)^{-3/2} $ is the maximum spin frequency at stellar surface. If
the critical fastness parameter is $\omega_{c}=0.55$, then the
maximum spin frequency can be as high as 715 Hz, which is similar
to the  fastest spin frequency of MSP observed (716 Hz). The lower
the critical fastness, the lower the final spin frequency
achieved.

For our numerical calculation, we take the effect of fastness
parameter into consideration.
If the critical fastness parameter is set as $0.9$, the numerical
solutions for equations (\ref{bt}), (\ref{pdot}), (\ref{omega})
and (\ref{ns}) suggest that the evolutionary curves in the $B-P$
diagram are a little below the spin-period equilibrium line.
However, the effect of the fastness parameter is significant  when
the B-P evolutionary track is close to the equilibrium period line
(spin-up line). This indicates that the spin angular velocity of
the star tends to reach the Keplerian angular velocity at the
inner edge of the accretion disk, but never exceeds this angular
velocity.  In fact, the primary  effect of the fastness parameter
is to force the evolution curves back to or below the equilibrium
period line and ensure that the evolutionary track cannot go
beyond this line.

From these tracks, we can see that despite of the differences for
initial fields and initial spin periods, the NSs can approach  at a
certain minimum spin period and bottom field after accreting $\sim >
0.2 \ms$. If the accreting phase is over, a recycled pulsar can be
found in the region of $B-P$ diagram with $B \sim 10^{8-9} $ G and
$P \sim 10 $ ms, forming a MSP. On account of the existence of
minimum spin-period and bottom field (e. g. $B \sim 10^{8} $ G), the
P-distribution and B-distribution imply that the MSPs accumulate
close to the region of low field and short period, which is an
explanation of the bimodal distributions for all pulsars, as shown
in Fig.\ref{his}. In short, for LMXBs the different initials give
the similar final value of spin-period, which is insensitive to the
initial conditions.

\subsection{Influence by Accretion Rate}

\begin{figure}
\includegraphics[width=9cm]{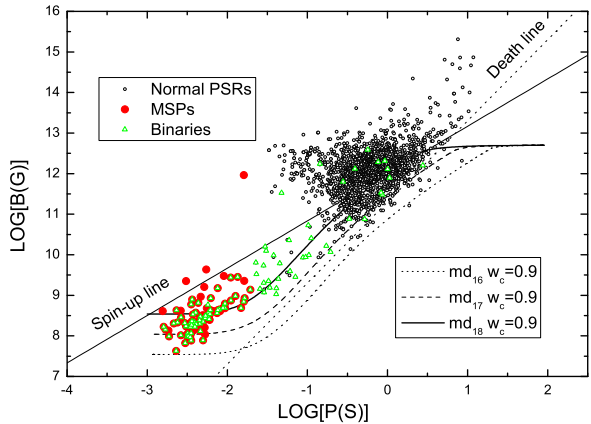}
\caption{The same meaning as Fig. \ref{B0} with different accretion
rates,   at the fixed initial magnetic-field ($\bo = 5 \times
10^{12} G$) and initial spin-period ($\po=100 s$), from the upper to
bottom, $\dot{M_{18}}$ - solid line, $\dot{M_{17}}$ - Dash line and
$\dot{M_{16}}$ - dot line, corresponding to the bottom fields of
$\sim 10^{7.5}$ G,  $\sim 10^{8}$ G and $\sim 10^{8.5}$ G,
respectively. The critical fastness parameter is set to be
$\omega_{c}$ = 0.9. } \label{MD}
\end{figure}

In this part, we study the influence by accretion rates on the
spin-period evolution. We set the accretion rate as $\dot{M_{16}},
\dot{M_{17}}$ and $\dot{M_{18}}$ at the fixed magnetic-field ($5
\times 10^{12}$ G) and spin-period ($100$ s), respectively(see Fig.
\ref{MD}). From the Fig. \ref{MD}, we find that, with different
accretion rate, the spin evolution follows different tracks. If we
set the critical fastness parameter as $0.9$, the tracks with
Eddington accretion rate ($\dot M_{18} = 1$) matches the evolution
of most recycled pulsars well, but the track with $\dot M_{16}$ lies
below the "death-line".   Moreover, the curves with a high accretion
rate and high critical fastness have the tendency to approach the
spin-up line.

\begin{figure*}

\centering$\begin{array}{c@{\hspace{0.19in}}c@{\hspace{0.19in}}c}
\multicolumn{1}{l}{\mbox{}} &
\multicolumn{1}{l}{\mbox{}} \\
\includegraphics[width=8cm]{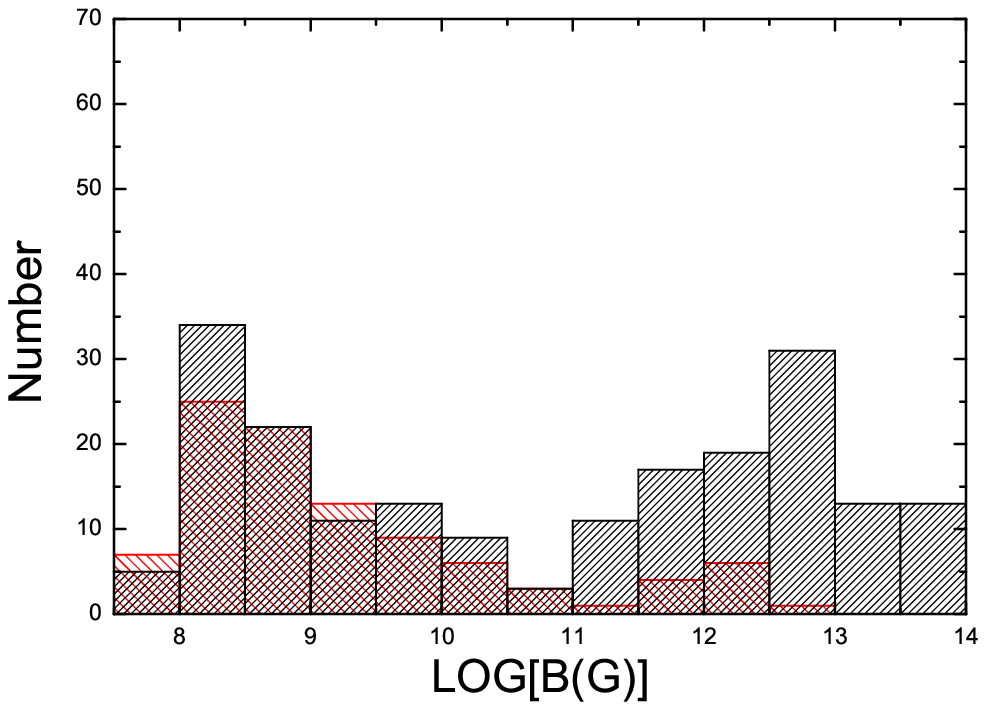} &\includegraphics[width=8cm]{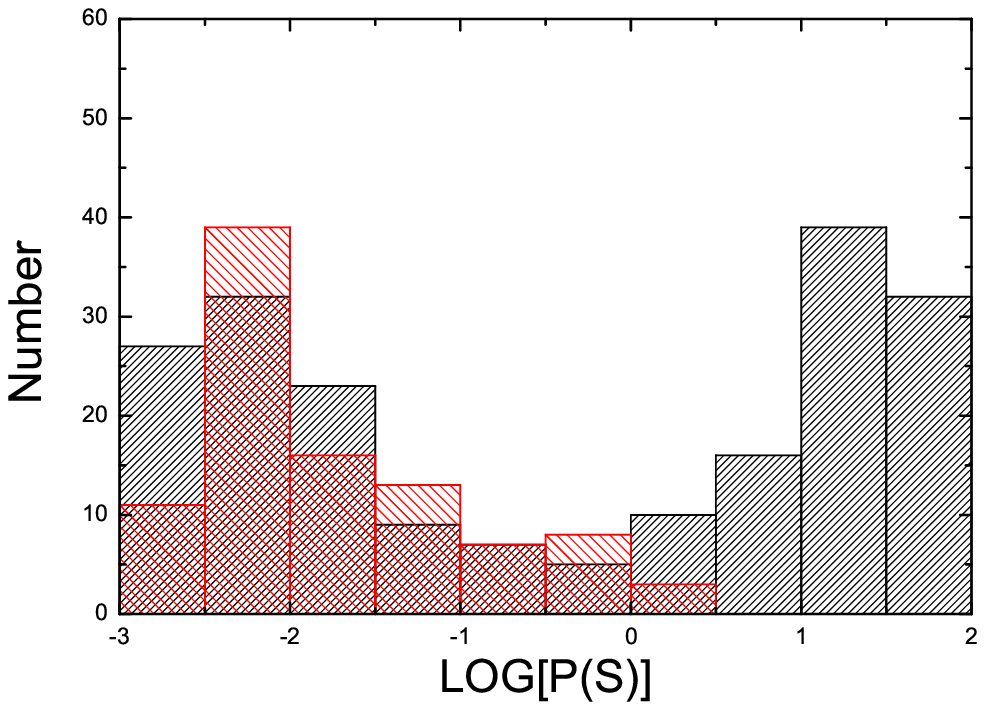}\\
\mbox{\bf (b)} & \mbox{\bf (c)}
\end{array}$
\caption{ Plots of histograms for spin-period and magnetic-field
obtained from the calculation for the theoretical model and  for the
observed binary pulsars (red-shaded).   In theoretical calculations,
we assume that the initial B-field follows a Gaussian distribution,
centered at $5\times10^{12}$ G, and ranged from $5\times10^{10}$G
(e.g. Hartman et al. 1997) to $10^{14}$G (e.g. Kaspi 2010); As for
the initial spin period, it is assumed to be Gaussian distribution,
centered at 30 s, ranged from 0.5 s to 100 s; The
 different accretion rates are set from  $\dot{M_{16}}$ to $\dot{M_{18}}$.
 The  theoretically evolved B-field and spin period values occurred at the left
 parts of plots, which are similar to the observed distributions.}\label{th-his}
\end{figure*}

{\bf          For a comparison of the magnetic-field and spin-period
distributions between theory and observation, we plot the
statistical distributions of magnetic-field and spin-period obtained
from the calculation for theoretical model with various initial
conditions (see caption of Fig. \ref{th-his}).
  100 samples of final B-field and final spin period (B, P ) have been done,
   which are plotted to compare the observed data of binary pulsars as
   shown  in Fig. \ref{th-his}, where we notice a bimodal distributions of
magnetic-field and spin-period of recycled pulsars (initial
distributions of B and P at right parts of plots compared to the
final evolved values at the left parts of plots). As compared to the
observational distributions of binary pulsars, we notice that the
both histograms for the MSPs (left parts of Fig. \ref{th-his}),
theoretical and observed data, are very similar. For the spin
period, more theoretical data than the observed values occurs at the
regimes close to millisecond, which may imply that more MSPs may not
evolve to their minimum spin period of about millisecond. }

\subsection{Spin-period Versus Accretion Mass ($P-\Delta M$)}

\begin{figure*}
\centering $\begin{array}{c@{\hspace{0.2in}}c@{\hspace{0.2in}}c}
\multicolumn{1}{l}{\mbox{}} & \multicolumn{1}{l}{\mbox{}} &
\multicolumn{1}{l}{\mbox{}} \\
\includegraphics[width=5.5cm]{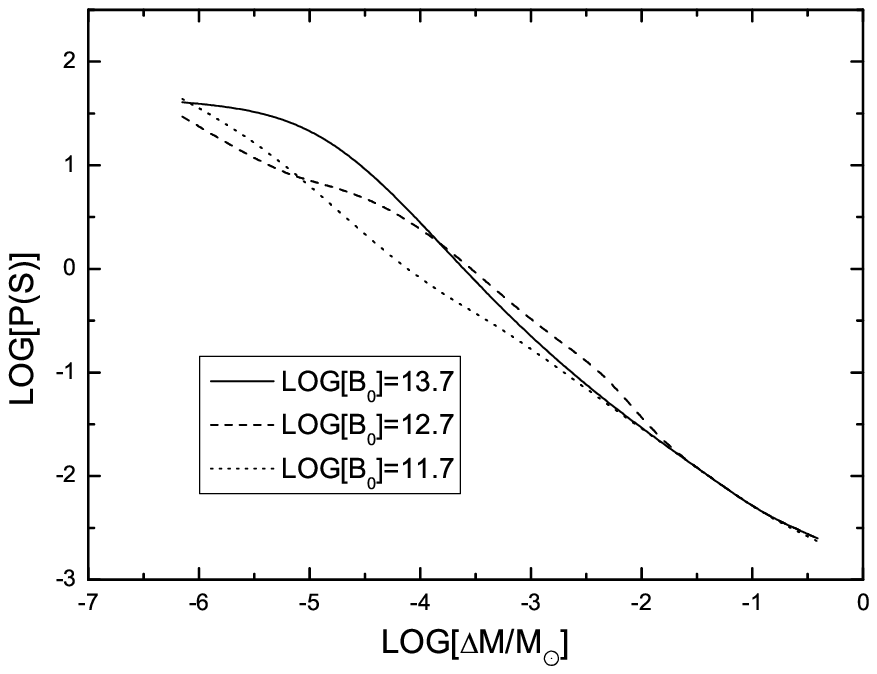} &\includegraphics[width=5.5cm]{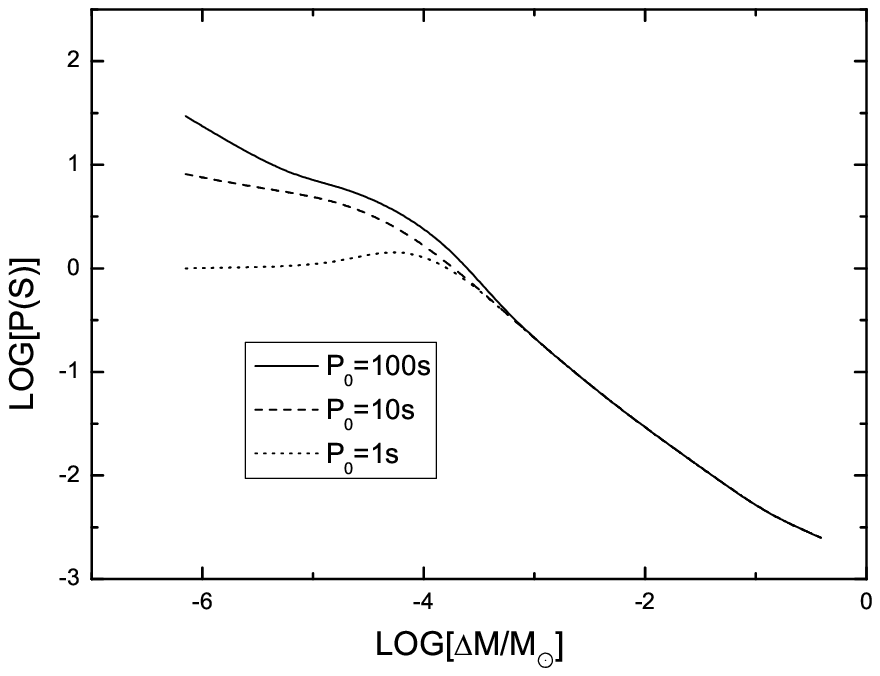}
&\includegraphics[width=5.5cm]{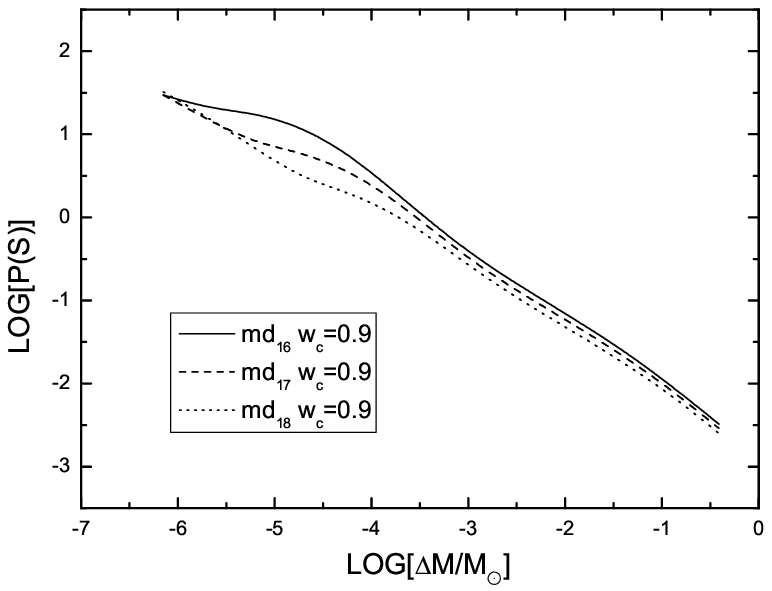}\\
\end{array}$
\caption{Plots of spin-period versus accretion mass with different
conditions, i.e., initial magnetic-fields (left), initial
spin-periods (middle) and accretion rates (right), where the
critical fastness parameter is set to be $\omega_{c} = 0.9$ in
calculation. It is noticed that the initial conditions ($B_{0}$ and
$P_{0}$) and accretion rate has little influence on the $P-\dm$ plot
after  the system accretes $\sim > 0.01 \ms$; In other words, the
spin periods  of  NS/LMXBs should be primarily related to how much
mass accreted.}\label{delm}
\end{figure*}

To investigate the spin-period evolution with accretion mass, we
consider the influences by initial conditions and accretion rate and
plot the evolutionary curves in $P-\Delta M$ diagrams (see Fig.
\ref{delm}). As a detailed illustration, Fig. \ref{delm} implies the
following results. (1). After accreting certain mass (e.g. $\sim
0.01\ms$), the spin-period evolution seems independent of the
initial magnetic-field. (2). The influence on $P-\Delta M$ by the
initial spin-period on the spin evolution exists when $\Delta M <
\sim 0.001 M_{\odot}$ and disappears when the accreted mass$\Delta M
> 0.001 M_{\odot}$. When a small amount of mass
($\sim < 0.001 M_\odot$) is transferred,   the spin-period mildly
changes (Fig. \ref{delm}), which may yield a HMXB with the
spin-period of some seconds, such as Her X-1 (van der Meer et al.
2007) and Vela X-1 (Quaintrell et al. 2003).  After a conveniently
long time, with the weight of material accreted $\sim 0.01 M_\odot$,
NS is spun-up to a shorter period ($P \sim < 100 ms$) which
corresponds to the system of DNS, e.g. PSR 1913+16 (Breton 2009).
The binaries with longer accretion phase, e.g.  LMXBs,  will accrete
sufficient mass ($\Delta M \sim > 0.1 M\odot$) from their companions
and yield a lower magnetic-field and shorter spin-period ($P \sim <
20$ ms), as in the case of MSPs, e.g. SAX J 1808.4-3658 (Wijnands \&
van der Klis 1998; Wijnands 2005) and PSR J 1748-2446 (Lorimer,
2008).  (3). At the beginning of accretion, NSs with different
accretion rates follow different curves in the $P-\dm$ plane. After
accreting $\sim 0.01 \ms$, the spin-period approximately follow the
same track with different accretion rate, until to the same minimum
period. Therefore, NSs in LMXBs, for the Z source with Eddington
luminosity and Atoll source with less luminosity, e.g. $\sim$ 1\%
Eddington accretion rate (Hasinger \& van der Klis 1989), should
have the similar spin frequencies while accreting $\sim 0.2\ms$.

\section{Summary and Discussion}
\label{sec4}

Considering the influences exerted by the different initial
magnetic-field, initial spin-period and the accretion rate, we
investigate the evolutionary tracks of recycled pulsar in $B-P$
diagram. The main results are listed below.

(1). From the evolutionary tracks in $B-P$ diagram, we find that the
magnetic-field of NS decays with accreting material, and the spin-up
evolution proceeds at the same time. The primary effect of the
fastness parameter is to force the evolutionary curves to reside
below the equilibrium period line. The critical fastness
($\omega_{c}$) can infer the final spin-period of recycled NS
(maximum spin frequency), for instance, $\omega_{c} = 0.5$  will
correspond to a spin of 650 Hz (1.54 ms), which is near the highest
observed value of MSP and AMXP.

(2). The influences of the initial magnetic-field, spin-period and
accretion rate are tested, respectively. The evolutionary curves
present that the spin-period (frequency) decreases (increases)
with the accretion and the minimum spin-period is  insensitive to
these two initial conditions when it is spun-up to $P \sim 10 $ ms
($\nu \sim 100  Hz$) after NS accreting about $\sim > 0.2
M_\odot$.

(3). It is found from the B-P diagram, a few observed MSPs locate
above the "spin-up line", while the trajectories based on our model
always lie below this line. Firstly, owing to the Ohmic dissipation,
the buried field of MSP may re-emerge after accretion, which can
account for a slightly increase of field of the star (e.g. Young \&
Chanmugam 1995; Bhattacharya 2008).
 Secondly, the magnetic model we use here is based on the "dilution of
polar magnetic flux" due to accretion, which is idealized. Due to
numerous plasma instabilities,  the Rayleigh-Taylor instability at
the disk-magnetosphere interface because of the high-density disc
matter being supported against by the low-density magnetosphere
plasma and Kelvin-Helmholtz instability because of the
discontinuity in the angular velocity of matter at the boundary
(Ghosh \& Lamb 1979a),  these instabilities may result in the
penetration of the magnetosphere, prying the field lines aside and
azimuthally wrapping the field lines by the disc matter (e.g.
Romanova 2008; Kulkarni \& Romanova et al. 2008),  which may
modify the field strength evolution then perturb the spin
evolution.

(4). The formation of the fastest spin frequency of MSP is an
interesting topic (Lorimer 2008; Chakrabarty et al. 2003;
Chakrabarty 2005). The known fastest radio pulsar is PSR
J1748-2446ad (Ter 5) at 716 Hz (Hessels et al. 2006), whereas the
highest AMXP  spin frequency in  LMXBs is 619 Hz (e.g. Wijnands
2005).  However, the recent report of 1122 Hz burst oscillation
frequency, interpreted as a spin frequency,  in XTE J 1739-285
(Kaaret et al. 2007) has been reported but not yet  confirmed on
account of the statistical significance (actual significance is only
$3\sigma$). Now the detected spin is less than the believed breaking
spin frequency of pulsar of about $\sim$ 1000 Hz (Lattimer \&
Prakash 2004), then what causes the special maximum spin frequency
of MSP?

The spin evolution is affected by the accretion torque,
which consists of two parts (Ghosh 1995; Frank et al. 2002): the
stresses associated with matter accreting from the inner edge of the
disk ($N_0 \equiv \dot{M}(GMR_{M})^{\frac{1}{2}}$) and that
associated with the magnetic field coupling the star with the disk.
The total torque $N$ can be conveniently expressed in terms of $N_0$
and the dimensionless torque. The spin-down torques which balances
the spin-up torque will also contribute to forming the maximum spin
frequency. Slow rotators ($\omega_s < \omega_c$) are spun up ($n >
0$) by this accretion torque. $\omega_s <  1$ is presumed, since for
$\omega_s > 1$ the steady accretion is of course impossible.  Here,
$\omega_c$ of approximately $\sim 0.3 - 0.5$ is the critical
fastness at which the torque changes direction, thus the maximum
spin frequency $\nu_{s} = 2\pi\Omega_{s} = \omega_{c}\nu_{k} \sim
(0.3 - 0.5) \nu_{k}$, where $\nu_{k} = 2\pi\Omega_{k}$. The observed
maximum upper kHz QPO frequency is 1330 Hz (van der Klis 2000; Zhang
et al. 2006) that is interpreted as the Keplerian frequency $\nk$ of
inner disk orbit, so this implies  a maximum spin frequency to be
about $\sim 400 - 665$ Hz which is consistent with observations.
%
%
To limit the occurrence of high spin of MSP over 1000 Hz,  the
braking torques have been considered.  Electromagnetic braking
torque can brake the stellar spin (Ghosh 1995, 2006), which  are
usually expressed in the form $n \propto \mu^2(\Omega_s/c)^3$,
where $\mu$ is the magnetic dipole moment of the star.
In NS/LMXB, the mass quadrupole can occur on account of the
accretion that arises the gravitational wave (GW) radiation
(Bildsten 1998; Melatos 2007; Melatos \& Payne 2005; Melatos \&
Peralta 2010; Ghosh 2006; Vigelius, Payne \& Melatos 2008; Watts et
al. 2009). The angular momentum is lost from the emission of GW
which may spin-down the NS.
With accretion, the gravitational wave and accretion torques
balance, but the total torque is effectively negative because the
gravitational wave torque increases with accreting. Hence, there
might arise a maximum spin frequency of several hundred Hertz for
a MSP.


\section{acknowledgements}

We are grateful for  the  discussions with G. Hasinger, M. Mendez,
and T. Belloni. This work is supported by the National Natural
Science Foundation of China (NSFC 10773017) and the National Basic
Research Program of China (2009CB824800).


\begin{thebibliography}{10}

\bibitem{}

\bibitem{}
Archibald A. M., Stairs I. H. \& Ransom S. M. et al. 2009, Science,
324, 1411

\bibitem{}
Alpar M. A., Cheng A. F. \&  Ruderman M. A. et al. 1982, Nature,
300, 728


\bibitem{}
Bhattacharya D. \& van den Heuvel E. P. J. 1991, Phys. Rep., 203, 1

\bibitem{}
Bhattacharya D. \& Srinivasan G. 1995, in X-ray Binaries, eds. Lewin
W. H. G., van Paradijs J. and van den Heuvel E. P. J., (Cambridge
University Press)



 \bibitem{}   Bhattacharya D. 2008, to appear in
Proceedings of "A Decade of Accreting Millisecond X-ray Pulsars",
eds. R. Wijnands et al.


\bibitem{}
Bildsten L. 1998, ApJ., 501, L89

\bibitem{}
Breton R. P. 2009, PhD thesis, arXiv:0907.2623





\bibitem{}
Burderi L., D'Antona F., Menna M. T., di Salvo T., Robba N. 2002a,
Mem. Soc. Astron. Ital., 73, 1072

\bibitem{} Burderi L., D'Antona F. \& Burgay M. 2002b, ApJ., 574, 325




\bibitem{}  Camilo F., Thorsett S. E. \& Kulkarni S. R. 1994, ApJ., 421, L15

\bibitem {}
Chakrabarty D., Morgan E.~H., Muno M.~P. et al. 
2003, Nature, 424, 42

\bibitem{}
Chakrabarty D. 2005, ASP Conf. Ser., 328, 279


\bibitem{}
Cheng  K. S. \& Zhang, C. M. 1998, A\& A,  337, 441


\bibitem{}
Cheng  K. S. \& Zhang, C. M. 2000, A\& A, 361, 1001



\bibitem{}
Cumming A., Zweibel E.~G. \& Bildsten L. 2001, \apj., 557, 958

\bibitem{}
Cumming A. 2005, ASP Conf. Ser., 328, 311


\bibitem{} Ferrario L. \& Wickramasinghe D. T. 2007a, AIP
Conference, 968, 188

\bibitem{} Ferrario L. \& Wickramasinghe D. T. 2007b, AIP
Conference, 968, 194


\bibitem{}
Elsner R. F., \& Lamb F. K. 1977, ApJ., 215, 897

\bibitem{}
Francischelli G. J., Wijers R. A. M. J. \& Brown G. E. 2002, ApJ.,
565, 471


\bibitem{}
Frank J., King, A. \& Raine D. J. 2002, Accretion Power in
Astrophysics, Cambridge, UK


\bibitem{}  Geppert U. \& Urpin V. 1994, MNRAS, 271, 490

\bibitem{}  Geppert U., Page D. \& Zannias T. 1999, \aap, 345, 847



\bibitem{} Ghosh P. \& Lamb F.K. 1977, ApJ., 217, 578

\bibitem{} Ghosh P. \& Lamb F.K. 1979a, ApJ., 232, 259

\bibitem{} Ghosh P. \& Lamb F.K. 1979b, ApJ., 234, 296

\bibitem{}
Ghosh, P. \& Lamb, F. K. 1992, in X-ray Binaries and Recycled
Pulsars, ed. E. P. J. van den Heuvel, S. A. Rappaport (Dordrecht:
Kluwer), 487

\bibitem{} Ghosh P. 1995, JApA, 16, 289

\bibitem{} Ghosh P. 2006, Rotation and Accretion Powered
Pulsars, World Scientific, India


\bibitem{}  Hartman J.W. et al. 1997, 325, 1031

\bibitem{}
Hasinger G. \& van der Klis M. 1989, \aap, 225, 79

\bibitem{}
Halpern, J. P. \& Gotthelf, E. F. 2010, ApJ., 709, 436

\bibitem{}
Hessels J. W., Ransom S. M. \& Stairs I. H. et al.
2006, Science, 311, 1901

\bibitem{}
Hurley J. R., Tout C. A., Wickramasinghe D. T., Ferrario, L. \& Kiel
P. D. 2009, MNRAS, 402, 1437 

\bibitem{}
Hobbs G. \& Manchester R. 2004, ATNF Pulsar Catalogue,
 http://www.atnf.csiro.au/research/pulsar/psrcat/psrcat\_help.html

\bibitem{}
Inogamov N. A. \& Sunyaev R. A., 1999, AstL., 25, 269





\bibitem{}
Kaaret P.,  Prieskorn Z. \& in't Zand J. et al. 2007, ApJ., 657, L97






\bibitem{} Kaspi V. M. 2010, PNAS, 107, 7147, arXiv:1005.0876v1

\bibitem{}
Konar S. ~\& Bhattacharya D.\ 1999, \mnras, 303, 588; 308, 795

\bibitem{} Konar S., \&
Choudhury A. 2004, \mnras, 348,  661


\bibitem{}
Kouveliotou C.  2003, AAS, 203, 7501

\bibitem{} Kulkarni A. K., \& Romanova M. M. 2008, MNRAS, 386, 673

\bibitem[Lamb \& Yu(2005)]{lamb-yu2005} Lamb F.~K. \& Yu W. 2005,
in Spin rates and magnetic fields of millisecond pulsars, in
Binary Radio Pulsars, ed.\ F. A. Rasio \&
Stairs I. H.(ASP Conference Series, Vol. 328), 299

\bibitem{}
Lattimer J. M. \& Prakash M. 2004, Science, 304, 536

\bibitem{} Li X. D., Wang Z. R., 1996, A\&A, 307, L5


\bibitem{}
Li X. D. \& Wang Z. R. 1999, ApJ., 513, 845

\bibitem{}
Lorimer D. R. 2008, Living Rev. Relativity, 11, 8
http://relativity.livingreviews.org/Articles/lrr-2008-8/

\bibitem{}
Lovelace R. V.,  Romanova M. M. \&  Bisnovatyi-Kogan G.S. 2005,
ApJ., 625, 957 

\bibitem{}
Lyne A. G., Burgay M., Kramer M. et al.
 2004, Science, 303, 1153 

\bibitem{}
Manchester R. N., Hobbs G. B., Teoh A. \& Hobbs M. 2005, AJ,
129, 1993 

\bibitem{}
Melatos A. 2007, AdSpR, 40, 1472

\bibitem{}
Melatos A. \& Phinney E. S. 2001, PASA, 18, 421

\bibitem{}
Melatos A. \& Payne D. 2005, ApJ., 623, 1044

\bibitem{}
Melatos A. \& Peralta C. 2010, ApJ., 709, 77


\bibitem{} Payne D. \& Melatos A. 2004, \mnras, 351, 569



\bibitem{Q03} Quaintrell H., Norton A. J. \&  Ash T. D. C. et al. 2003, A\&A, 401, 313

\bibitem{}   Radhakrishnan V. \& Srinivasan G.  1982,  Curr.
Science,  51,  1096



\bibitem{} Romani G. M., 1990, Nature, 347, 741


\bibitem{}
Romanova M. M., Kulkarni A. K. \& Lovelace R. V. E. 2008, ApJ., 673,
L171

\bibitem{}
Shapiro S. L. \& Teukolsky S. A. 1983, Black Holes, White Dwarfs and
Neutron Stars. Wiley, New York

\bibitem{}
 Shibazaki N., Murakami T., Shaham J. \& Nomoto K. 1989,
 Nature, 342, 656



\bibitem{} Taam R. E. \& van den Heuvel E. P. J. 1986, ApJ., 305, 235

\bibitem{} Urpin V. \& Geppert U. 1995, MNRAS, 275, 1117

\bibitem{} Urpin V. \& Konenkov D. 1997, 284, 741


\bibitem{} Urpin V., Geppert U. \& Konenkov D. 1997, MNRAS,
295, 907

\bibitem{} Urpin V., Geppert U. \& Konenkov D. 1998, A\&A,
331, 244


\bibitem{}  van den Heuvel E. P. J.  1995, JA\&A, 16, 255

\bibitem{}  van den Heuvel E. P. J. \& Bitzaraki O. 1994, MmSAI,
65, 237

\bibitem{}  van den Heuvel E. P. J. \& Bitzaraki O. 1995a, A\&A, 297, L41

\bibitem{}  van den Heuvel E. P. J. \& Bitzaraki O. 1995b, In:
The Lives of the Neutron Stars, Kluwer Academic Publishers,
Dordrecht

\bibitem{}
van den Heuvel E. P. J., 2004, science, 303, 1143

\bibitem{}
van der Klis M. 2000, ARA\&A, 38, 717 


\bibitem{vanderMeer:2007hj}
van der Meer A., Kaper L., van Kerkwijk M. H., Heemskerk M. H. M. \&
van den Heuvel E. P. J. 2007, A\&A, 473, 523



\bibitem{}
Vigelius M., Payne D., Melatos A. 2008, Proceedings of the 11th
Marcel Grossmann Meeting on General Relativity, World Scientific,
arxiv: 0811.2031

\bibitem{} Watts A.L. et al. 2009, MNRAS, 389, 839

\bibitem{}
Wickramasinghe D. T., Hurley J. R., Ferrario1 L., Tout C. A. \& Kiel
P. D. 2009, JPhCS., 172, 2037

\bibitem{}
Wijnands R. \& van der Klis M. 1998, Nature, 394, 344



\bibitem{}
Wijnands, R. 2006, Trends in Pulsar Research, ed. J. A. Lowry (New
York, USA: Science Publishers, Inc.), 53



\bibitem{}
Wijers R. A. M. J. 1997, \mnras, 287, 607


\bibitem{}
Yin H.X., Zhang C.M. \& Zhao Y.H. 2007, A\&A, 471, 381,
arXiv:0705.1431



 \bibitem{}
 Young E.J.  \& Chanmugam G.  1995, ApJ, 442, L53


\bibitem{}
Zhang C. M. \& Kojima Y. 2006, MNRAS, 366, 137

\bibitem{} Zhang  C. M., Yin  H. X., \& Zhao  Y. H., et al. 2006,
 \mnras, 366, 1373.


\end{thebibliography}
\end{document}